\documentclass[conference, anonymous]{IEEEtran}
\IEEEoverridecommandlockouts
\usepackage{cite}
\usepackage{amsmath,amssymb,amsfonts}
\usepackage{algorithmic}
\usepackage{textcomp}
\usepackage{xcolor}
\usepackage{algorithm}
\usepackage{url}
\usepackage{svg}
\usepackage{booktabs} 
\usepackage{graphicx} 
\usepackage{multirow} 
\def\BibTeX{{\rm B\kern-.05em{\sc i\kern-.025em b}\kern-.08em
    T\kern-.1667em\lower.7ex\hbox{E}\kern-.125emX}}
\begin{document}

\title{eACGM: Non-instrumented Performance Tracing and Anomaly Detection towards Machine Learning Systems\\
\thanks{\textsuperscript{\dag}Ruilin Xu and Zongxuan Xie contributed equally to this work.}
\thanks{979-8-3315-4940-4/25/\$31.00~\copyright~2025 IEEE}
}

\author{
Ruilin Xu\textsuperscript{\dag}, Zongxuan Xie\textsuperscript{\dag}, Pengfei Chen \\
School of Computer Science and Engineering, Sun Yat-sen University, Guangzhou, China \\
\{xurlin5, xiezx25\}@mail2.sysu.edu.cn, chenpf7@mail.sysu.edu.cn
}

\maketitle

\begin{abstract}
We present \textbf{eACGM}, a full-stack AI/ML system monitoring framework based on eBPF. eACGM collects real-time performance data from key hardware components, including the GPU and network communication layer, as well as from key software stacks such as CUDA, Python, and PyTorch, all without requiring any code instrumentation or modifications. Additionally, it leverages \texttt{libnvml} to gather process-level GPU resource usage information. By applying a Gaussian Mixture Model (GMM) to the collected multidimensional performance metrics for statistical modeling and clustering analysis, eACGM effectively identifies complex failure modes, such as latency anomalies, hardware failures, and communication inefficiencies, enabling rapid diagnosis of system bottlenecks and abnormal behaviors.

To evaluate eACGM's effectiveness and practicality, we conducted extensive empirical studies and case analyses in multi-node distributed training scenarios. The results demonstrate that eACGM, while maintaining a non-intrusive and low-overhead profile, successfully captures critical performance anomalies during model training and inference. Its stable anomaly detection performance and comprehensive monitoring capabilities validate its applicability and scalability in real-world production environments, providing strong support for performance optimization and fault diagnosis in large-scale AI/ML systems.
\end{abstract}


\section{Introduction}

The scale and complexity of modern AI/ML systems continue to grow, with extensive GPU computation and low-level communication playing crucial roles~\cite{ref1, ref2}. In multi-GPU and multi-node training, resource contention and complex scheduling often lead to performance issues, such as CUDA memory overflows, kernel timeouts, and GPU contention~\cite{ref7, ref8}, which can slow down or even interrupt training. Real-time detection and localization of such failures are thus vital for system stability and efficiency.

Existing monitoring tools primarily rely on instrumentation~\cite{ref9, ref10, ref31}, which, despite improving observability, introduce high overhead and may interfere with training, especially in distributed settings. Moreover, they often fail to offer real-time, full-stack monitoring across hardware and software layers. The dynamic nature of GPU scheduling and heterogeneous stacks further increases monitoring complexity. In long-running workloads, conventional techniques struggle to pinpoint bottlenecks and failures promptly, extending diagnosis cycles.

To address this, an efficient, low-overhead, full-stack monitoring solution is needed. Recently, extended Berkeley Packet Filter (eBPF) has emerged as a lightweight, high-efficiency monitoring technology. Running in the kernel, eBPF collects critical interaction data in real time without modifying application code, making it suitable for complex AI/ML environments.

In this paper, we propose \textbf{eACGM}, an \textbf{e}BPF-based \textbf{A}utomated \textbf{C}omprehensive \textbf{G}overnance and \textbf{M}onitoring framework. eACGM offers full-stack monitoring from hardware (GPU, network) to software (CUDA, Python, PyTorch) with minimal system impact. It integrates \texttt{libnvml}~\cite{ref34} for fine-grained GPU metrics and applies GMM-based analysis to identify anomalies and performance bottlenecks, supporting rapid fault localization and system tuning.

Our key contributions are summarized as follows:

\begin{enumerate}
    \item \textbf{Non-intrusive, real-time monitoring}: eACGM uses eBPF for low-overhead tracking of key performance metrics, enabling near “zero-intrusion” monitoring.
    \item \textbf{Fine-grained GPU tracking}: With \texttt{libnvml}, eACGM monitors process-level utilization, memory, temperature, and power.
    \item \textbf{Full-stack observability}: eACGM covers both hardware (GPU, network) and software (CUDA, Python, PyTorch) layers.
    \item \textbf{Intelligent anomaly analysis}: Using GMM, eACGM models high-dimensional metrics to detect anomalies and guide optimization.
\end{enumerate}

By introducing eACGM, we significantly enhance AI/ML system monitoring capabilities, providing a novel approach to performance optimization and fault diagnosis. This work lays the foundation for future advancements in large-scale AI/ML system observability, offering promising prospects for practical deployment. The source code of eACGM is available at \url{https://github.com/shady1543/eACGM}.

\begin{figure*}[t]
\centering
\includegraphics[width=1\textwidth]{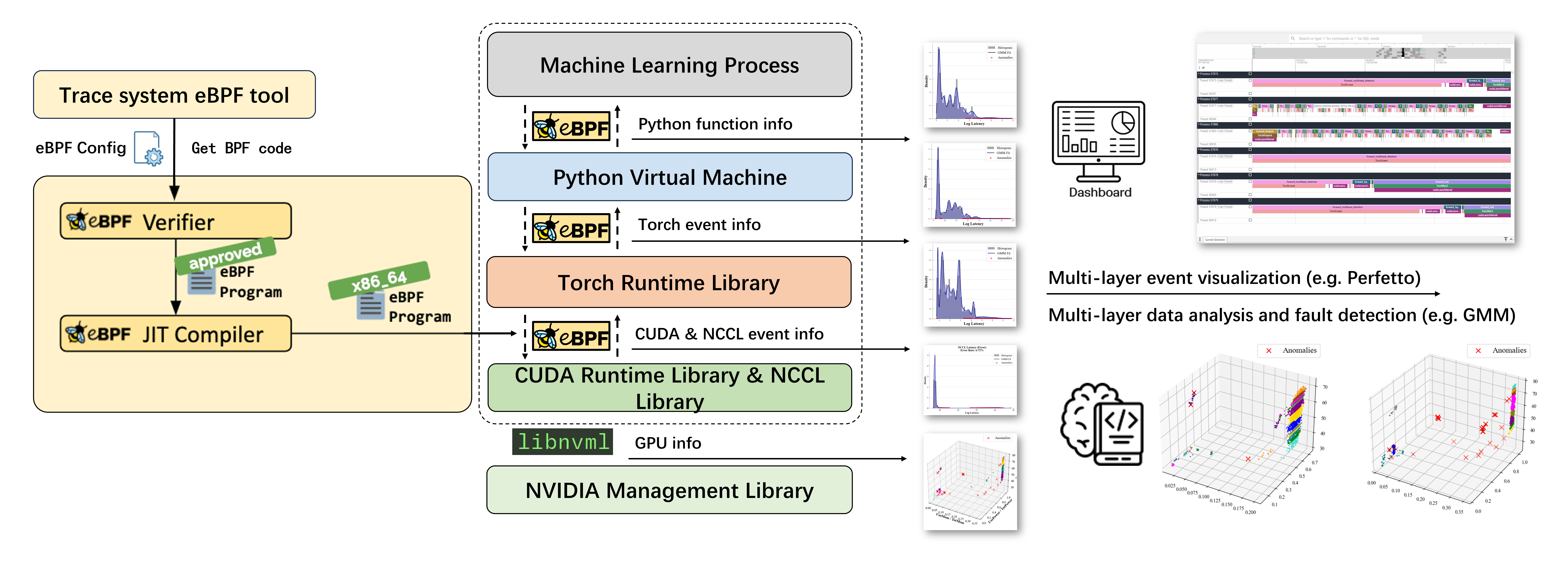}
\caption{The proposed eACGM architecture, enabling full-stack monitoring across software (CUDA, Python, PyTorch) and hardware (GPU, NCCL) layers using eBPF and \texttt{libnvml}.}
\label{fig:arch}
\end{figure*}

\section{Related Work}

\subsection{AI/ML Performance Profiling and Diagnosis}  
Existing AI/ML analysis tools mainly rely on instrumentation or explicit API calls. Tools like Nvprof and Nsight Systems~\cite{ref9,ref10} analyze GPU kernel statistics to aid CUDA optimization, but require code changes or command-line configurations, introducing overhead. Framework-level profilers such as PyTorch and TensorFlow Profilers~\cite{ref11,ref12} provide user-friendly APIs, yet depend on manual instrumentation and cannot transparently capture cross-layer bottlenecks or low-level communication dynamics.

\subsection{GPU Resource Monitoring and Analysis}  
Tools like DCGM and Nvidia-smi~\cite{ref13,ref14} monitor GPU utilization, power, and temperature in real time. While widely used in cluster diagnostics, they focus on hardware state visualization and lack integration with AI/ML frameworks. As a result, they provide limited insight into how framework-level events correlate with hardware anomalies, restricting their utility in full-stack diagnosis.

\subsection{eBPF-based System Monitoring}  
Initially developed for network analysis~\cite{ref18,ref19}, eBPF now supports observability, security, and debugging~\cite{ref20,ref21,ref24}. Tools like bcc, bpftrace, and Sysdig~\cite{ref15,ref16,ref17} attach kernel-space probes for real-time, low-overhead state capture. However, existing tools focus on general system tracing and lack AI/ML-specific support. In particular, they do not capture GPU usage, framework events, or distributed communication metrics critical to AI/ML workloads.

\subsection{Distributed Training and Communication Optimization}  
Distributed training introduces communication latency and resource contention. Prior work improves efficiency via NCCL tuning, workload scheduling, and topology planning~\cite{ref37,ref41,ref42}, but lacks full-stack, real-time monitoring to detect communication anomalies and correlate them with system events. As AI/ML scales across nodes and GPUs, dynamic bottlenecks emerge that require integrated observability.

To address these gaps, \textbf{eACGM} uses eBPF to trace system events without instrumentation and integrates \texttt{libnvml} for fine-grained GPU metrics. Unlike traditional tools, eACGM correlates low-level kernel events with high-level AI frameworks, using GMM to detect anomalies across the full stack, enabling non-intrusive, intelligent monitoring for large-scale AI/ML workloads.

\section{System Design and Implementation}

\subsection{System Architecture}

eACGM is a full-stack monitoring framework for AI/ML systems. It leverages eBPF for kernel-level event tracing with low overhead and integrates \texttt{libnvml} to capture detailed GPU performance data.

As shown in Figure~\ref{fig:arch}, eACGM spans from low-level CUDA events to high-level PyTorch and Python operations. eBPF probes are dynamically inserted at key execution points to trace function calls, kernel launches, and operator executions. Meanwhile, \texttt{libnvml} provides GPU metrics such as utilization, memory, and power. Collected data can be visualized via tools like Perfetto~\cite{ref48} and analyzed using GMM for fault detection and performance diagnosis.

\subsection{Data Collection}

eACGM combines eBPF probes and \texttt{libnvml} queries to gather fine-grained performance data across software and hardware layers.

\textbf{Tracing CUDA Events.}  
eACGM identifies key CUDA functions from the PyTorch runtime and system path, placing eBPF probes to trace memory allocation and kernel launches. This reveals bottlenecks such as memory inefficiencies or kernel timeouts.

\textbf{Tracing Python Calls.}  
eBPF attaches to \texttt{PyObject\_CallFunction} to monitor Python calls with timestamps and thread IDs, helping identify overhead from frequent invocations or blocking operations.

\textbf{Tracing Torch Operators.}  
Despite C++ symbol obfuscation, eACGM locates relevant PyTorch runtime functions via reverse engineering, enabling operator-level tracking. This supports profiling pre- and post-JIT acceleration and detecting operator bottlenecks.

\textbf{Tracing NCCL Events.}  
eACGM instruments NCCL APIs (e.g., \texttt{ncclAllReduce}) to measure latency and message size, uncovering communication bottlenecks in distributed setups.

\textbf{Process-level GPU Monitoring.}  
Using \texttt{libnvml}, eACGM captures per-process GPU metrics (memory, utilization), aiding in diagnosing contention and imbalance.

\textbf{Global GPU Monitoring.}  
It also tracks overall GPU metrics (e.g., power, temperature) to detect large-scale anomalies and ensure system stability.

This layered data collection builds a holistic profile for AI/ML workloads, supporting effective diagnosis and optimization.

\subsection{Data Analysis}

eACGM correlates multi-source data (CUDA, Python, Torch, NCCL, GPU) to uncover bottlenecks and inefficiencies.

\textbf{CUDA Event Analysis.}  
It examines kernel configurations and memory allocation patterns to identify suboptimal launch settings or memory fragmentation.

\textbf{Python Call Analysis.}  
By profiling call frequency and duration, eACGM detects overhead from repetitive or blocking Python calls.

\textbf{Torch Operator Analysis.}  
eACGM measures operator runtimes (e.g., \texttt{TorchLinear}, \texttt{TorchConv2d}), supporting analyses such as JIT effects or performance bottlenecks.

\textbf{NCCL Communication Analysis.}  
By analyzing NCCL event latency and message size, it guides optimization of distributed training communication.

\textbf{GPU Utilization Analysis.}  
It correlates memory, power, and temperature trends to detect imbalance and contention, improving resource efficiency.

Through multi-layer analysis, eACGM delivers actionable insights for optimizing performance and enhancing AI/ML system stability.

\section{Fault and Performance Bottleneck Analysis}

\subsection{Statistical Modeling}

Inspired by~\cite{ref26, ref29}, we adopt a statistical modeling approach for observability in AI/ML systems. Under consistent conditions, system events and performance metrics exhibit stable statistical patterns. We model these patterns using a Gaussian Mixture Model (GMM), which clusters system states based on feature distributions (e.g., latency, resource usage). This forms the basis for unsupervised fault diagnosis and performance bottleneck detection~\cite{ref28}.

A GMM models the dataset \(X = \{x_1, \dots, x_N\}\) as a mixture of \(K\) Gaussian components:

\[
p(x) = \sum_{k=1}^{K} \pi_k \mathcal{N}(x | \mu_k, \Sigma_k)
\]

where \(\pi_k\) is the weight of the \(k\)-th component, and \(\mathcal{N}(x | \mu_k, \Sigma_k)\) denotes the multivariate normal distribution.

The parameters $\{\pi_k, \mu_k, \Sigma_k\}$ are estimated using the Expectation-Maximization (EM) algorithm, a widely used method for fitting Gaussian Mixture Models, as shown in Algorithm~\ref{alg:EM}.

\begin{algorithm}[ht]
\caption{Expectation-Maximization (EM) Algorithm for GMM}
\label{alg:EM}
\begin{algorithmic}[1]
\STATE \textbf{Input:} Dataset \(X = \{x_1, \dots, x_N\}\), number of components \(K\)
\STATE \textbf{Output:} Estimated parameters \(\{\pi_k, \mu_k, \Sigma_k\}\)
\STATE Initialize parameters randomly
\REPEAT
    \FOR{each \(x_i\)}
        \STATE Compute responsibility \(\gamma(z_{ik})\)
    \ENDFOR
    \FOR{each component \(k\)}
        \STATE Update \(\pi_k, \mu_k, \Sigma_k\)
    \ENDFOR
\UNTIL{convergence}
\end{algorithmic}
\end{algorithm}

eACGM trains a GMM over recent data (e.g., past hour) using features such as CUDA calls, Torch operators, GPU usage, and communication latency. For each new event, it computes the probability of belonging to each component. If all probabilities fall below a threshold, the event is flagged as anomalous. This approach leverages the GMM’s ability to model multimodal behavior and separate normal and abnormal states.

\subsection{Fault Detection and Bottleneck Identification}

Anomaly detection is critical for identifying deviations from normal operation, such as latency spikes or resource inefficiencies. eACGM uses the trained GMM to probabilistically classify system states and detect anomalies.

\vspace{0.1cm}
\noindent
\textbf{Definition 1.} \textit{(Anomaly Detection Criterion). An event \(x_i\) is flagged as anomalous if its probability density under the most likely component is below a threshold \(\delta\):}
\[
p(x_i | \theta_k) < \delta
\]
\textit{where}
\[
p(x_i | \theta_k) = \frac{1}{\sqrt{(2\pi)^d |\Sigma_k|}} \exp\left(-\frac{1}{2} (x_i - \mu_k)^T \Sigma_k^{-1} (x_i - \mu_k)\right)
\]

\begin{figure*}[h]
\centering
\includegraphics[width=1\textwidth]{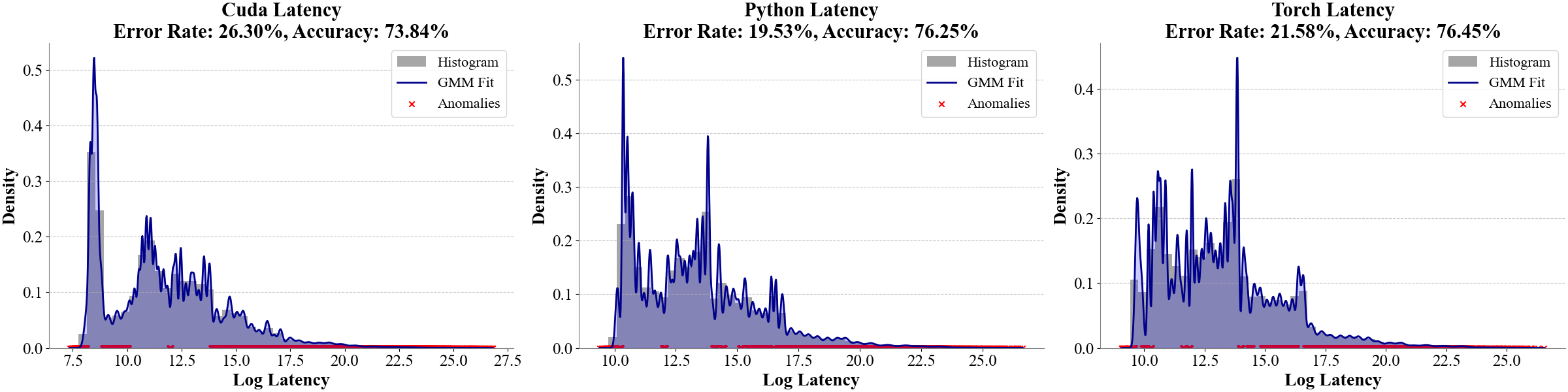}
\caption{Latency anomaly detection with eACGM.}
\label{fig:Latency}
\end{figure*}

The anomaly detection procedure assigns each event to the most probable component and computes its density. Events falling below \(\delta\) are considered outliers. The algorithm is outlined in Algorithm~\ref{alg:AnomalyDetection}.

\begin{algorithm}[ht]
\caption{Anomaly Detection Algorithm}
\label{alg:AnomalyDetection}
\begin{algorithmic}[1]
\STATE \textbf{Input:} GMM parameters \(\{\pi_k, \mu_k, \Sigma_k\}\), dataset \(X = \{x_1, \dots, x_N\}\), threshold \(\delta\)
\STATE \textbf{Output:} Anomalous events \(A\)
\STATE \(A \gets []\)
\FOR{each \(x_i \in X\)}
    \STATE Find \(k = \arg \max_k p(x_i | \theta_k)\)
    \STATE Compute \(p(x_i | \theta_k)\)
    \IF{\(p(x_i | \theta_k) < \delta\)}
        \STATE Add \(x_i\) to \(A\)
    \ENDIF
\ENDFOR
\RETURN \(A\)
\end{algorithmic}
\end{algorithm}

This statistical approach enables eACGM to detect potential faults and bottlenecks by identifying deviations in event behavior, providing a robust and quantitative basis for performance diagnosis.

\section{Experiments}

\subsection{Experimental Setup}

We evaluate eACGM on a multi-node, multi-GPU GPT-2 training task. Experiments are conducted on a dual-node cluster, each equipped with an Intel Xeon Gold 6326 CPU @2.90GHz, 128GB RAM, six A40-48GB GPUs, and a ConnectX-6 NIC, representing realistic compute- and communication-intensive AI workloads.

To quantitatively assess detection accuracy, we construct a labeled dataset by injecting faults during training. The dataset contains over 1M samples, with a normal-to-anomalous ratio of approximately 5:1, simulating the typical imbalance in real-world system monitoring. Each data point includes features such as CUDA and Torch events, GPU metrics, and communication latencies.

\subsection{Latency Anomaly Detection}

Latency issues in distributed AI workloads can stem from scheduling inefficiencies, operator delays, or hardware-level slowdowns. We inject latency-related faults at multiple layers to emulate such issues:

\textbf{Software Faults.}  
Using \texttt{pytorchfi}~\cite{ref32}, we introduce artificial delays into matrix multiplications and activation functions to simulate inefficient operator behavior.

\textbf{CUDA Faults.}  
Via \texttt{DCGM}~\cite{ref33}, we simulate kernel timeouts and memory errors to induce CUDA-level latency.

eACGM traces latency at the CUDA, Python, and PyTorch layers using eBPF, then applies GMM to identify anomalies. The detection accuracies reach 73.84\%, 76.25\%, and 76.45\% at the respective layers. Fig.~\ref{fig:Latency} shows the detection results, where red crosses indicate identified anomalies.

\subsection{Hardware Anomaly Detection}

\begin{figure}[h]
\centering
\includegraphics[width=0.35\textwidth]{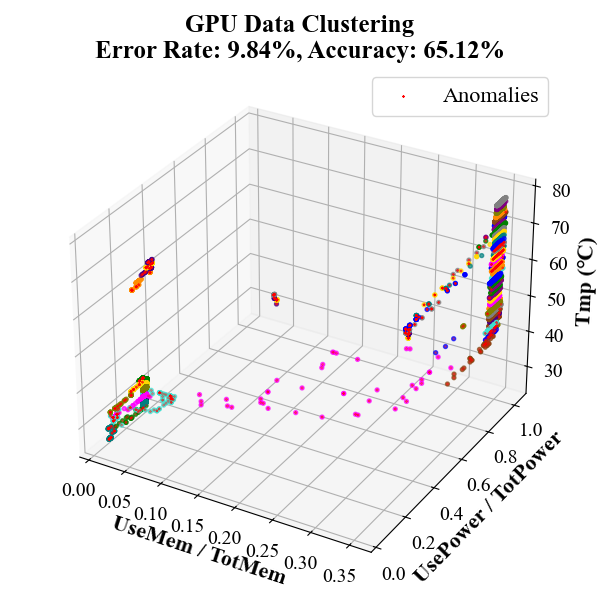}
\caption{Hardware anomaly detection with eACGM.}
\label{fig:GPU}
\end{figure}

We simulate resource contention by mapping multiple processes to shared GPUs, causing abnormal memory, power, and utilization patterns. eACGM uses \texttt{libnvml}~\cite{ref34} to monitor GPU metrics (e.g., utilization, memory, temperature for illustration) and applies GMM clustering for anomaly detection, achieving 65.12\% accuracy. Fig.~\ref{fig:GPU} visualizes the results, where the pink outliers indicate detected anomalies.

\begin{table*}[h]
\centering
\caption{Comparison of accuracy, recall, and F1 scores across various methods.}
\begin{tabular}{l c c c c c c c c c c c }
\toprule
\textbf{Model}                          &                           & \textbf{KMeans} & \textbf{Isolation Forest} & \textbf{DBSCAN} & \textbf{XGBoost}  & \textbf{SVM}  & \textbf{Random Forest}      & \textbf{GMM}     \\ \midrule
\multirow{5}{*}{\textbf{Accuracy}}      & \textbf{Latency (CUDA)}   & 62.10\%         & 61.38\%                   & 60.45\%         & 69.02\%           & 68.30\%       & 70.24\%                     & \textbf{73.84\%} \\
                                        & \textbf{Latency (Python)} & 61.57\%         & 66.32\%                   & 65.17\%         & 69.87\%           & 67.15\%       & 71.04\%                     & \textbf{76.25\%} \\
                                        & \textbf{Latency (Torch)}  & 62.98\%         & 68.42\%                   & 66.01\%         & 71.10\%           & 69.43\%       & 73.58\%                     & \textbf{76.45\%} \\
                                        & \textbf{Hardware}         & 55.24\%         & 61.15\%                   & 58.17\%         & 62.40\%           & 61.22\%       & 64.34\%                     & \textbf{65.12\%} \\
                                        & \textbf{NCCL}             & 64.79\%         & 70.45\%                   & 69.16\%         & 73.26\%           & 72.11\%       & 75.00\%                     & \textbf{85.04\%} \\ \midrule
\multirow{5}{*}{\textbf{Recall}}        & \textbf{Latency (CUDA)}   & 59.73\%         & 58.12\%                   & 57.83\%         & 63.04\%           & 61.90\%       & 64.13\%                     & \textbf{73.89\%} \\
                                        & \textbf{Latency (Python)} & 58.13\%         & 63.45\%                   & 60.21\%         & 62.23\%           & 61.10\%       & 63.94\%                     & \textbf{75.63\%} \\
                                        & \textbf{Latency (Torch)}  & 58.88\%         & 63.80\%                   & 61.42\%         & 64.99\%           & 63.13\%       & 66.35\%                     & \textbf{78.17\%} \\
                                        & \textbf{Hardware}         & 52.50\%         & 55.02\%                   & 56.56\%         & 58.89\%           & 57.78\%       & 54.98\%                     & \textbf{59.52\%} \\
                                        & \textbf{NCCL}             & 61.56\%         & 68.22\%                   & 64.45\%         & 69.34\%           & 68.11\%       & 71.56\%                     & \textbf{80.07\%} \\ \midrule
\multirow{5}{*}{\textbf{F1}}            & \textbf{Latency (CUDA)}   & 60.45\%         & 59.12\%                   & 58.64\%         & 63.06\%           & 61.88\%       & 64.11\%                     & \textbf{75.00\%} \\
                                        & \textbf{Latency (Python)} & 59.11\%         & 65.04\%                   & 61.88\%         & 64.12\%           & 62.45\%       & 65.12\%                     & \textbf{74.12\%} \\
                                        & \textbf{Latency (Torch)}  & 60.09\%         & 64.55\%                   & 62.50\%         & 66.08\%           & 64.45\%       & 67.21\%                     & \textbf{72.57\%} \\
                                        & \textbf{Hardware}         & 55.22\%         & 54.03\%                   & 56.23\%         & 51.12\%           & 56.54\%       & 52.23\%                     & \textbf{58.73\%} \\
                                        & \textbf{NCCL}             & 64.01\%         & 69.34\%                   & 66.19\%         & 69.02\%           & 67.55\%       & 71.23\%                     & \textbf{80.80\%} \\ \bottomrule
\end{tabular}
\vspace{0.1cm}
\label{tab:acc_comparison}
\end{table*}

\subsection{Communication Anomaly Detection}

\begin{figure}[h]
    \centering
    \includegraphics[width=0.45\textwidth]{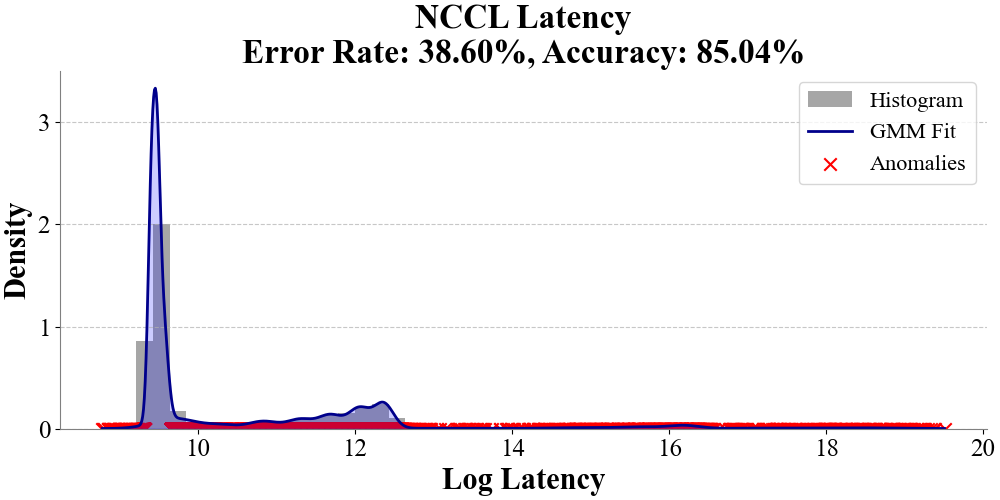}
    \caption{Communication anomaly detection with eACGM.}
    \label{fig:NCCL}
\end{figure}

To demonstrate eACGM’s capability in communication monitoring, we use \texttt{chaosblade}~\cite{ref43} to inject network faults, including latency and packet loss. eACGM traces NCCL events and applies GMM to communication-level metrics such as message latency and bandwidth. Using latency as a representative example, eACGM achieves 85.04\% detection accuracy. Fig.~\ref{fig:NCCL} shows the results, where red crosses indicate identified anomalies.

\subsection{Comparison with Other Monitoring Tools}

We compare eACGM with several mainstream tools in the GPT-2 training setting:

\begin{itemize}
    \item \textbf{cProfile}~\cite{ref44}: Profiles Python only; no GPU/CUDA support.
    \item \textbf{Torch Profiler}~\cite{ref11}: Covers Python and CUDA; requires code changes.
    \item \textbf{NCCL Trace}~\cite{ref46}: Limited to NCCL layer tracing.
\end{itemize}

Unlike these tools, eACGM is zero-intrusive and supports full-stack monitoring across CUDA, Python, Torch, and NCCL layers. Powered by eBPF and \texttt{libnvml}, it enables comprehensive system analysis without code modifications. Table~\ref{tab:comparison} summarizes the comparison.

\begin{table}[ht]
\centering
\caption{Comparison of eACGM with other monitoring tools.}
\label{tab:comparison}
\begin{tabular}{ccc}
\toprule
\textbf{Tool} & \textbf{Monitored Layer(s)} & \textbf{Invasive} \\
\midrule
cProfile      & Python                    & No  \\
Torch Profiler & Python, CUDA             & Yes \\
NCCL Trace    & NCCL                     & No  \\
\textbf{eACGM} & CUDA, Python, Torch, NCCL & No  \\
\bottomrule
\end{tabular}
\vspace{0.1cm}
\end{table}

\subsection{Comparison with Other Clustering Methods}

As shown in Table~\ref{tab:acc_comparison}, GMM achieves the best results in terms of accuracy, recall, and F1-score on all monitored layers. Notably, GMM leads in recall, indicating strong anomaly detection capability, and maintains the highest F1-scores on NCCL and Torch layers. In addition, GMM delivers stable and superior accuracy across CUDA, Python, Torch, hardware, and NCCL data, reflecting its robustness in diverse system scenarios. Compared to KMeans~\cite{ref49}, Isolation Forest~\cite{ref50}, DBSCAN~\cite{ref51}, XGBoost~\cite{ref52}, SVM~\cite{ref53}, and Random Forest~\cite{ref54}, GMM consistently outperforms both traditional clustering and supervised methods, confirming its effectiveness for full-stack system monitoring.









\subsection{Sensitivity Analysis}

\begin{figure}[H]
\centering
\includegraphics[width=0.473\textwidth]{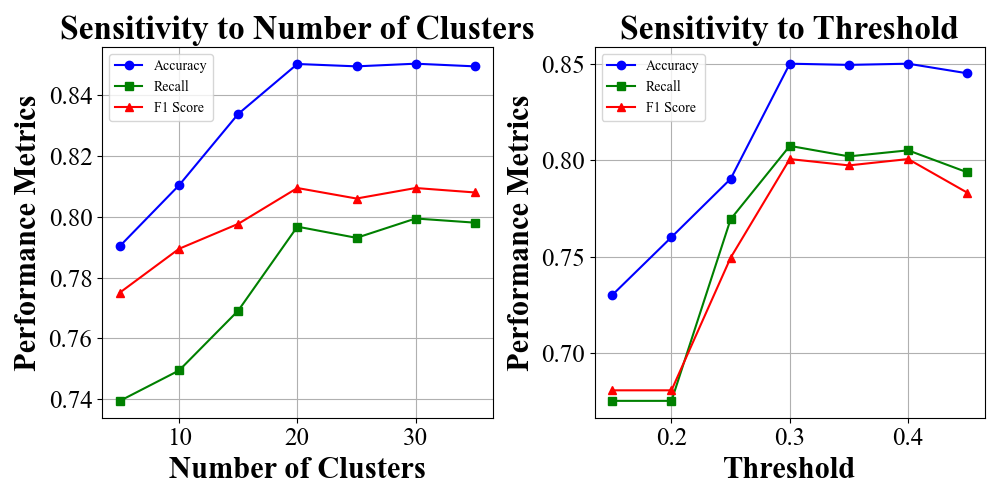}
\caption{Sensitivity analysis of GMM.}
\label{fig:sen}
\end{figure}

We analyze GMM sensitivity to the number of components and threshold \(\delta\), using NCCL latency data. As shown in Fig.~\ref{fig:sen}, results are stable under parameter variations, though overly small values degrade accuracy.

\section{Conclusion}

We introduced \textbf{eACGM}, an eBPF-based framework for full-stack monitoring of AI/ML systems in multi-node, multi-GPU environments. eACGM seamlessly integrates system metrics from the GPU, network, and application layers (including CUDA, Python, and PyTorch), and leverages \texttt{libnvml} for process-level GPU resource monitoring. By applying Gaussian Mixture Models (GMM) for quantitative clustering and anomaly detection, eACGM accurately identifies latency, hardware, and communication anomalies, enabling rapid fault localization and performance optimization. Experimental results validate that eACGM provides non-intrusive, full-stack monitoring and significantly enhances system reliability.

\section*{Acknowledgment}
The research is supported by National Key Research and Development Program of China under Grant 2024YFB4505904, the Guangdong Basic and Applied Basic Research Foundation (No.2023B1515020054). The corresponding author is Pengfei Chen.


\end{document}